\documentstyle{article}
\begin{document} 
\centerline{\bf A generalized spin model of}
\centerline{\bf financial markets}

\vspace{1cm}

\centerline{\bf Debashish Chowdhury\footnote{On leave from Physics Department, I.I.T., Kanpur 208016, India} and Dietrich Stauffer} 
\centerline{Institute for Theoretical Physics,}
\centerline{University of Cologne,}
\centerline{D-50923 K\"oln,} 
\centerline{Germany} 

\vspace{1cm}

\noindent{\bf Abstract:} We reformulate the Cont-Bouchaud model 
of financial markets in terms of classical "super-spins" where 
the spin value is a measure of the number of individual traders  
represented by a portfolio manager of an investment agency. We 
then extend this simplified model by switching on {\it interactions} 
among the super-spins to model the tendency of agencies getting 
influenced by the opinion of other managers. We also introduce a 
fictitious temperature (to model other random influences), and 
time-dependent local fields to model a slowly changing optimistic or 
pessimistic bias of traders. We point out close similarities between 
the price variations in our model with $N$ super-spins and total 
displacements in an $N$-step Levy flight. We demonstrate the 
phenomena of natural and artificially created bubbles and subsequent 
crashes as well as the occurrence of "fat tails" in the distributions 
of stock price variations. 

\vspace{2cm}


\newpage
\section{Introduction:}

The mathematical modelling of economic phenomena in stock- and 
currency-markets has been going on for one century [1-4]. However, 
recently physicists have begun applying the concepts and techniques 
of statistical physics to understand the dynamical behaviour of 
these "complex adaptive systems" by developing models which are 
similar, at least in spirit, to the statistical mechanical models 
of interacting microscopic constituents of macroscopic samples 
of matter. The constituent elements in these "microscopic" models 
of markets represent the individual investors and investment 
agencies [5-11]. 

Cont and Bouchaud (CB) ~\cite{cb} have suggested one of the 
simplest models of financial markets; this model has led to 
interesting conclusions regarding the "microscopic" origin of the 
"herd behaviour", "bubbles" and "crashes" at the stock markets. 
There exists a close relation between this theory and the theory 
of percolation ~\cite{stauaha} (see also ~\cite{staupen}; for a 
short review of microscopic models see ~\cite{stauap}). By 
simplifying the CB model through a reformulation and, then, 
extending it further, in this paper we develop a more detailed 
model of stock market; this is formulated in terms of {\it 
interacting super-spins}, which are maintained at a {\it 
fictitious temperature} and which evolve with time following a 
{\it stochastic dynamics}, in the presence of {\it time-dependent 
local fields}. We explain the motivations for these reformulations 
and extensions of the CB model and examine the corresponding 
consequences by analyzing the temporal fluctuations in the changes 
of stock prices.

\section{The Models:} 

\subsection{The CB Model:} 

In the CB model ~\cite{cb}, pairs of individual investors are 
linked randomly with probability $p$ and the clusters of linked 
individuals thus formed are identified as "coalitions" of 
investors; all the members of each coalition make the same 
investment decision (i.e., whether to buy or to sell or not to  
trade). Therefore, each cluster may correspond, for example, to 
funds managed by the same portfolio manager. 

Since, in the original formulation of this model ~\cite{cb}, a 
link is allowed to form between any pair of investors, the 
clustering corresponds to bond percolation in infinite-dimensional 
space ~\cite{stauaha}. Isolated individual investors may be viewed 
as clusters of size one. Once the individual investors form the 
clusters, the time-evolution of the clusters proceeds as follows: 
each cluster randomly decides to buy (with probability $a$), to 
sell (with probability $a$) or not to trade (with probabilty $1-2a$) 
during each unit time interval. The change of the stock price is 
defined to be proportional to the difference between the demand 
and supply.  If $n_s^+$ is the number (per investor) of the buying 
clusters and $n_s^-$ is the number (per investor) of the selling 
clusters then the price change $\Delta$ is given by 
$\Delta \propto [\sum_s s ~n_s^+ - \sum_s s ~n_s^-]$.

\subsection{The super-spin model:} 

In this paper we first reformulate the CB model in terms of 
"superspins". The system consists of super-spins $S_i$; the 
magnitude $|S_i|$ of the super-spins are drawn from a pre-determined 
probability distribution $P(|S|)$ and each spin can be in one of 
the three possible states, viz., $+|S_i|, 0, -|S_i|$. These 
superspins are analogues of the clusters in the CB model and 
the magnitude of the spin corresponds to the cluster size in 
the CB model. At every discrete time step, each of the super-spins 
choses the state $+|S_i|$ with probability $a$, the state $-|S_i|$ 
with probability $a$ and the state $0$ with probability $1-2a$; 
this is identical to the rule of time evolution of the clusters 
of investors in the CB model. The total number of {\it individual} 
investors is $\sum_{i=1}^N |S_i|$ and the price change, which is 
defined to be proportional to the difference in the total demand 
and total supply, is thus proportional to the total magnetization 
$M = \sum_{i=1}^{N} S_i$ where $N$ is the total number of 
super-spins (i.e., the total number of clusters of investors). 

\section{Results and Discussion:}

Using this reformulated version of the CB model, together with 
the distribution 
\begin{equation}
P(|S|) \propto |S|^{-(1+\alpha)}, 
\end{equation}
we have computed the distributions of stock price variations for 
several different values of $a$ and $N$. This distribution is 
non-Gaussian, irrespective of the number of investment agencies 
(see fig.1a; $\alpha=3/2$ in fig.1). 

The super-spin model, as formulated above, is closely related to  
{\it Levy flights}. By Levy flight one means a random walk where 
the the probability $p(\ell)$ of a jump of size $\ell$ is given by 
the distribution~\cite{weiss,hughes} 
\begin{equation}
p(\ell) \propto \ell^{-(1+\alpha)} \quad with \quad 0 < \alpha < 2. 
\end{equation}
Since the form (1) of $P(|S|)$ in our model is identical to that 
of $p(\ell)$ for a Levy flight, the magnetization of $N$ super-spins 
(i.e., the price change of the stocks) is the analogue of the total 
displacement after $N$ steps of a particle performing Levy-flights; 
this is similar to the concepts introduced originally by Mandelbrot 
~\cite{mandel} and also to the stochastic multiplicative process of
Levy and Solomon \cite{solomon}. Therefore, $P(M)$, the distribution of the 
stock price variations in our model may appear to be the distribution 
$P_{LF}^{(N)}(x)$ of the total displacements $x$ of $N$-step Levy flights. 
However, that is not true as there is a subtle difference between the 
two processes arising from the fact that, in our model, the spin 
configuration $\{S\}$ (analogue of the $N$ displacements of the Levy 
flight) is generated from another configuration by using the rule 
that a spin $S_i$ decides to be in the state $\pm |S_i|$ and $0$ with 
the probabilities $a$ and $1-2a$, respectively. Therefore~\cite{sornlevy}, 
if $n$ is the number of non-zero superspins in a configuration $\{S\}$, 
\begin{equation}
P(M) = \sum_{n=0}^{N} {N \choose n} (2a)^n (1-2a)^{N-n} P^{(n)}(M),   
\end{equation}
where $P^{(1)}(M)$ is identical to the distribution (1) for $P(|S|)$ 
and $P^{(n)}(M)$ represents the distribution obtained by $n$ convolutions 
of $P(|S|)$ with itself. Taking Fourier transform of both sides of (3) 
we get
\begin{equation}
{\hat P}(k) = \sum_{n=0}^{N} {N \choose n} (2a)^n (1-2a)^{N-n}
{\hat P}^{(n)}(k)   					
\end{equation}
where ${\hat P}(k)$ and ${\hat P}^{(n)}(k)$ are the Fourier 
transforms of $P(M)$ and $P^{(n)}(M)$, respectively. Now,
\begin{equation}
{\hat P}^{(n)}(k) = [{\hat P}^{(1)}(k)]^n			
\end{equation}
where ${\hat P}^{(1)}(k)$ is nothing but the Fourier transform of (1). 
Inserting (5) into (4) we get a series that can be summed analytically 
and, hence, 
\begin{equation}
{\hat P}(k) = [2a {\hat P}^{(1)}(k) + (1-2a)]^N      		
\end{equation}
The expression of ${\hat P}^{(1)}(k)$ is known exactly. For the 
simplicity of analysis we consider the case where $\alpha <1$. 
Then, for small $k$ (i.e., large price variations) 
\begin{equation} 
{\hat P}^{(1)}(k) = 1 - C |k|^{\alpha} + {\rm higher~ order~ terms}.
\end{equation}
Inserting (7) into (6) we get 
\begin{equation}
{\hat P}(k) = 1 - 2a N C |k|^{\alpha} + {\rm higher~ order~ terms}, 
\end{equation}
which implies that the tail of the distribution of $M$ has the same 
exponent $1+\alpha$ as our input in equation (1). Next, we consider 
the case where $ 1 < \alpha < 2$. In this case, (7) is replaced by 
\begin{equation}
{\hat P}^{(1)}(k) = 1 - ik g - C |k|^{\alpha} + {\rm higher~ order~ terms}.
\end{equation} 
Inserting (9) in to (6) now we get
\begin{equation}
P(k) = \exp [-2Na [ik g + C |k|^{\alpha} +  {\rm higher~ order~ terms}]]
\end{equation}
which also implies that the tail of the distribution of $M$ has the 
same exponent $1+\alpha$ as our input in equation (1). This is, indeed, 
consistent with our numerical data obtained from computer simulation 
(see fig.1(b)) and is trivial in the limit $a \rightarrow 0$. 

Moreover, from the above analysis one would expect the tail of the 
distribution of the stock price variations in our super-spin model to 
have the same exponent $1+\alpha$ as our input in equation (1) for 
all velues of $a$. This is, indeed, what we observed by replotting 
our data on a log-log plot (not shown in any figure) after scaling 
the widths of all the non-Gaussian distributions of fig.1c to unity.
In contrast to these features of $P(M)$ in our super-spin formulation, 
the distribution of price variations in the CB model is close to a 
Gaussian for sufficiently large $a$, at least when formulated on 
finite-dimensional lattices (see fig.1c) although it is non-Gaussian 
with a power-law tail for small $a$. 

Furthermore, equation (10) also suggests that, if the 
distribution $P(|S|)$ is given by the equation (1) then, in the 
asymptotic regime of large $M$, the {\it amplitude} of the tails in 
the distribution $P(M)$ should scale linearly with $a$ for small $a$ 
on a semi-log plot. 
The lack of good agreement between this theoretical prediction and our 
numerical data is most probably caused by the fact that the true 
asymptotic regime may be far beyond the largest $M$ plotted in this 
figure. It is also worth pointing out here that for small $k$ 
(i.e., for large variation of the prices) the dependence of $P(M)$ 
on $a$ and $N$ enters through the product $aN$ and this is consistent 
with our observation. 

In order to model the tendency of traders (individual as well 
as portfolio management agencies) getting influenced by the 
opinion of other traders, we now "switch on" {\it interactions} 
among  the super-spins. The super-spins are {\it not} located on 
the sites of any lattice. We define the {\it "total opinion "} 
$H_i$ gathered  by the $i$-th trader, because of all the other 
traders, as  
\begin{equation}
H_i = \sum_{j \neq i} J_{ij} S_j 
\end{equation} 
where $J_{ij}$ is a measure of the strength of the mutual influence 
between the pair of traders (individual or investment agencies) 
labelled by $i$ and $j$; the sum on the right hand side of (11) is 
to be carried out over all the $N-1$ super-spins excluding $i$. 
If $J_{ij} > 0$ ($J_{ij} < 0$), then a buying $j$-th trader would 
encourage the $i$-th trader to buy (sell), and vice versa. Note 
that, in the language of the spin models of magnetism, $J_{ij}$ is 
the strength of the exchange interaction between spin-pairs and 
$H_i$ is the Weiss molecular field (or, internal field). 

For our discussion here, we consider only the natural choice, 
namely, $J_{ij} > 0$ for all pairs $(ij)$. Moreover, for simplicity, 
we consider $J_{ij} = J$, a constant independent of $i$ and $j$, 
for all the spin-pairs, so that the "total opinion" gathered by 
the $i$-trader can be written as 
\begin{equation}
H_i = J \sum_{j \neq i} S_j .
\end{equation} 
Note that $H_i = 0$ describes a balance of optimistic and pessimistic 
traders whereas $H_i > 0$ ($H_i < 0$) correspond to predominant 
optimism (pessimism). The motivation for including not only the sign 
of $S_j$ ~\cite{lux, mark}, but also its magnitude on the right hand 
side of (12) comes from the fact that, usually, the larger is a trading 
agency the stronger is its effect on shaping the market opinion. 
Although, in real markets, this effect of $S_j$ may be nonlinear, 
i.e., not proportional to the first power of the size of the trading 
agency, we assume a linear dependence for the sake of simplicity. 

At this stage of formulation of our model, every trader may be 
regarded as a "noise trader" who has no own opinion  and 
would decide whether to buy or sell depending on whether the 
"total opinion" gathered is positive or negative. Besides, the 
larger is the magnitude of $H_i$ the stronger will be the 
corresponding opinion influencing the decision of the $i$-th 
"noise trader". We define the {\it "disagreement function"} $E_i$ 
of the $i$-th noise trader as 
\begin{equation}
E_i = - S_i H_i = - J \sum_{j \neq i} S_i S_j; 
\end{equation}
all "noise traders" would like to minimize the corresponding  
disagreement function. In the language of spin models of magnetism, 
$E_i$ is the energy of the $i$-th spin because of its interactions 
with the other spins. 

If all the spins minimized their energies the system of super-spins 
would end up in a ferromagnetic state. Equivalently, if all the 
traders mimimized their disagreement defined above, i.e., if all the 
investors make decision only depending on what other investors are 
doing (and minimize their own disagreement accordingly) the market 
will end up in either of two possible states where all the traders 
will either like to buy or sell. However, this does never happen in 
any real financial markets because the traders neither blindly follow 
the market opinion nor can always manage to follow the market opinion 
(even if they wanted to) because of so many reasons other than the 
influence of all other traders. In order to model these random 
influences, which are not explicitly included in our model, we introduce 
a fictitious temperature $T$. Since, in reality, the average price 
change usually vanishes, and since price change corresponds to the 
total magnetization in our model, we choose a sufficiently large 
magnitude of $T$ so that the magnetization fluctuates in time about 
the zero mean value. Therefore, we modify the dynamics of the model 
as follows: a super-spin picks up the states $+|S_i|, -|S_i|$ and $0$ 
with the probabilities $a, a$ and $1-2a$, respectively and, then it 
is allowed to make a transition from its current state to the 
state it has picked up with the probability $e^{-\Delta E_i/(k_BT)}$
where $\Delta E$ is the change in its energy (i.e., the change of 
disagreement in the language of economics) associated with this 
transition. 

Finally, we further extend this reformulated model to incorporate 
"fundamentalist traders" who form at least a part of their opinion 
(i.e., optimistic or pessimistic bias) towards the stocks of a 
company on the basis of an analysis of the fundamentals of that 
company. If $h_i$ is the "individual bias" of the $i$-th investor, 
then the corresponding "disagreement function" is given by 
\begin{equation}
E_i = - S_i (H_i + h_i); 
\end{equation}
where a positive $h_i$ corresponds to optimism while a negative 
$h_i$ indicates pessimism of the trader. The dependence of $h_i$ 
on $i$ implies that different fundamentalist trading agencies 
can have different evaluations of the fundamental value. However, 
in contrast to the time-independent local fields considered usually 
in spin models, both the magnitude as well as the sign of the 
"individual bias" $h_i$ of the traders can change with time. We 
study the effects of this time-dependent "individual bias" on the 
price variations. 

When every super-spin is subjected to a random local field which 
is positive and negative with equal probability but has the same 
magnitude, the system represents a market where every trading agent 
is a fundamentalist but the biased opinion of the agents happen to 
be randomly optimistic or pessimistic with equal probability. In 
such a situation we find that the fluctuations in the price variation 
can be much stronger even when $|h_i|$ is not too strong (merely 
comparable to $T$). Nevertheless, the qualitative nature of the 
price variations in a market where all the traders are fundamentalist 
is no different from that in a market where every trading agent is a 
"noise trader" so long as the fundamentalists have rigid opinions 
which do not change with time (see fig.2).  

We now consider a market where $50 \%$ of the trading agents 
are fundamentalists (all with very strong bias towards the stocks 
of the company under consideration) while the other trading agents 
are all noise traders. In other words, to begin with, each of the 
randomly chosen $50 \%$ of the agencies is subject to a 
positive local field of sufficiently high magnitude. In this case, 
we also impose the condition that when the price variation becomes 
too high or too low (i.e., cross a tolerance window) the fundamentalists 
reverse their bias. This is implemented in our super-spin model by 
"flipping" the direction of each of the local fields. Thus, the 
local fields switch from positive to negative when $M$ rises above 
a positive value which is chosen apriori, say $0.4$, and reverse 
switching from negative to positive local fields take place 
when $M$ falls below $-0.4$. The occurrence of larger values of 
$|M|$ in this situation (see fig.3) implies that the fundamentalist 
traders with apparently very strong optimistic bias can push up the 
demand for the stocks of a company, even if they represent a fraction 
of all traders, but when they reverse their opinion, it triggers a 
rush for selling the stocks. In this way the nearly periodic variation 
of fig.3b is produced, showing that this last variant of the model is
unrealistic. The 
qualitatively different distribution of price variations observed in 
such cases (fig.4) indicates that the statistics of bubbles and 
crashes created by the strong bias of a few fundamentalists would be 
very different from those of commonly encountered ones which are, 
thus, dominated by less rational behaviour.  

\section{Summary and conclusion:} 

In this paper we have developed a model of stock market which may 
be viewed as a model of interacting super-spins which are maintained 
at a fictitious temperature and are subjected to time-dependent 
local fields, where the stochastic dynamics of the super-spins 
describes the temporal evolution of the decisions (i.e., whether to 
buy, or sell or not to trade) of individual investors and investment 
agencies; this dynamics, in turn, leads to the temporal fluctuations 
of the stock price. We have studied the nature of these fluctuations 
of the stock price and the phenomena of bubbles and crashes. 

In the CB model the probability $p$, with which individual traders 
are linked to form clusters, is tuned to be identical with (or 
very close to) the corresponding percolation threshold thereby 
guaranteeing a power-law distribution of the cluster sizes which, 
in turn, leads to the desired behaviour of the price variations. 
In our model the distribution of the superspins is directly tuned 
to a power law. It would be interesting to develop a model which 
can "self-organize" so as to produce coalitions whose sizes are 
distributed according to the desired power law.

\vspace{1cm}

We thank D. Sornette and T. Lux for communicating useful informations 
and the Alexander von Humboldt Foundation for support.

\newpage

\noindent{\bf Figure Captions:} 

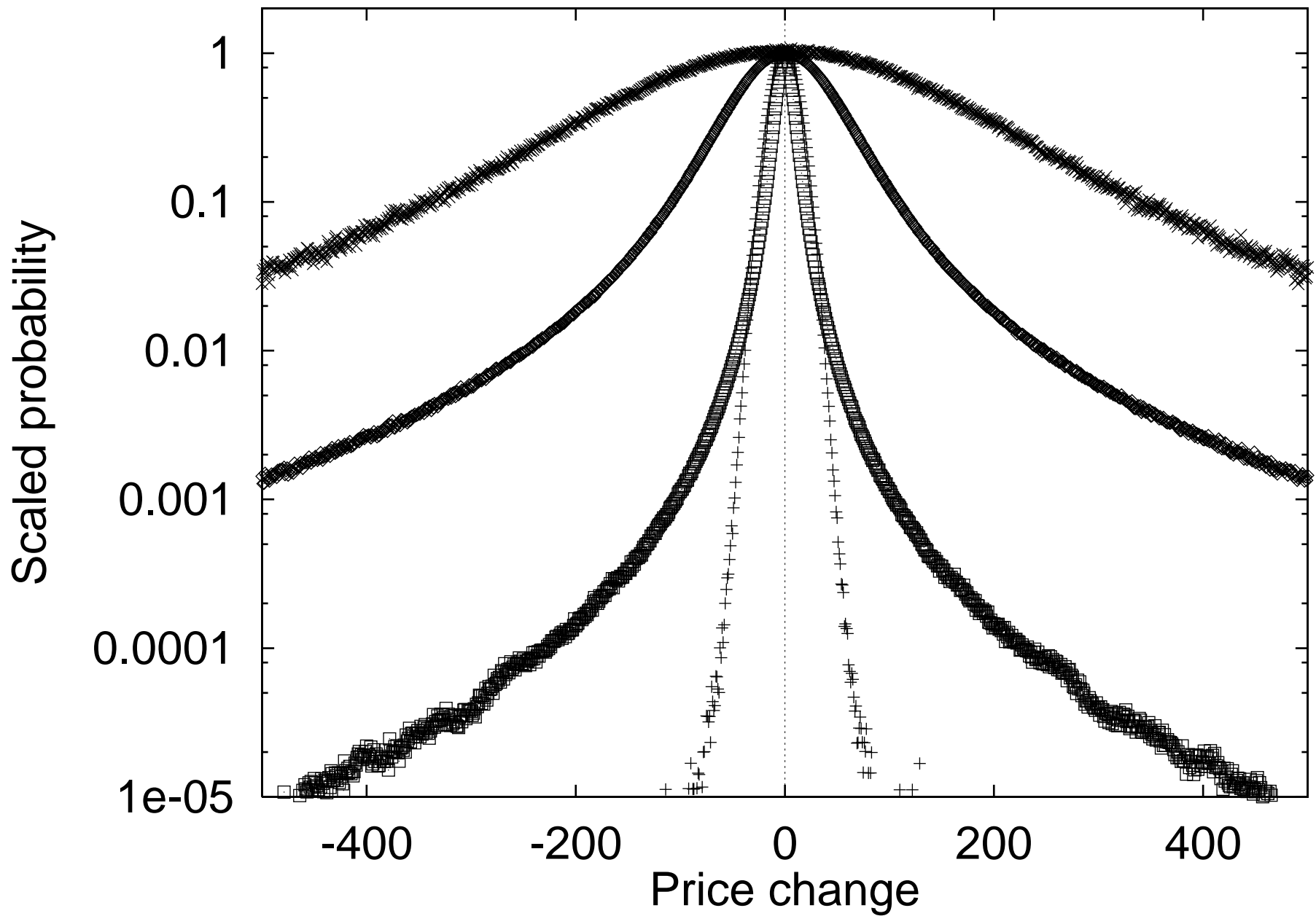
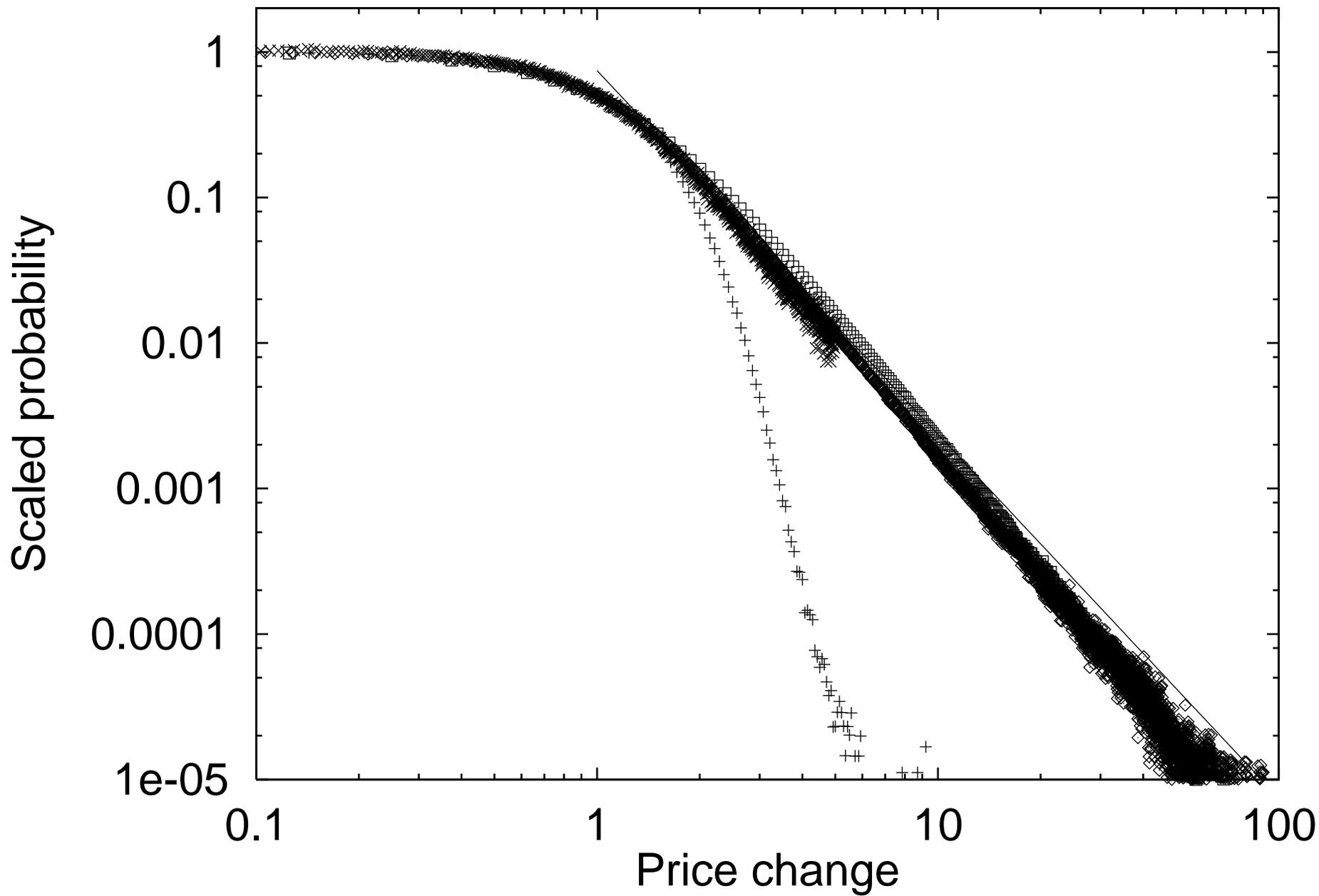
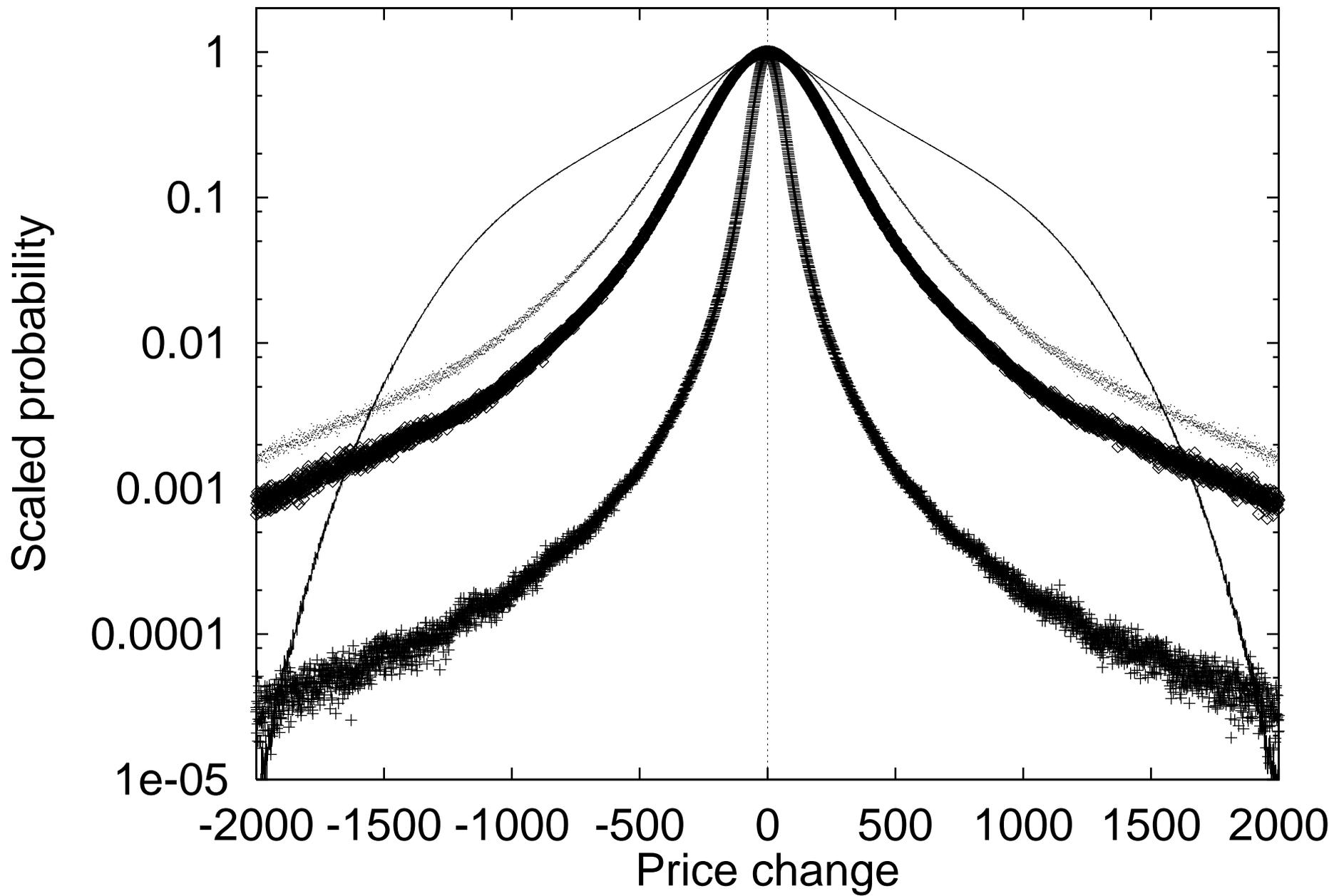
\noindent{\bf Fig.1:}\\
{\bf (a)} After rescaling the heights  of the 
distributions of the price changes to unity, the  
distributions are shown on a semi-log plot. The symbols 
$\Box, \diamond$ and $\times$ correspond to $N = 10^2, 10^3$ 
and $5000$, respectively (all for $\alpha = 3/2$) while the 
symbol $+$ correspond to $\alpha = 7/2, N = 10^3$. For all the 
curves, $a = 0.05$.\\
{\bf (b)} After rescaling the widths of the distributions in 
fig.1a also to unity, the distributions are shown on a log-log 
plot using the same symbols as in fig.1a. The input distribution 
(1), with $\alpha = 3/2$, has been represented by the straight 
line.\\
{\bf (c)} After rescaling the heights  of the 
distributions of the price changes to unity, the scaled 
distributions are shown on a semi-log plot. The symbols $\dots, 
\diamond$ and $+$ correspond to $a = 0.5, 0.33, 0.05$, respectively 
(all for $N = 1000$). For comparison, the data for a CB model 
system of $71 \times 71$ traders, run up to $1000$ 
iterations with $ a = 1/3$ and averaged over $10^6$ samples, 
are shown with a line similar to fig.3 of Stauffer and Penna \cite{staupen}. 

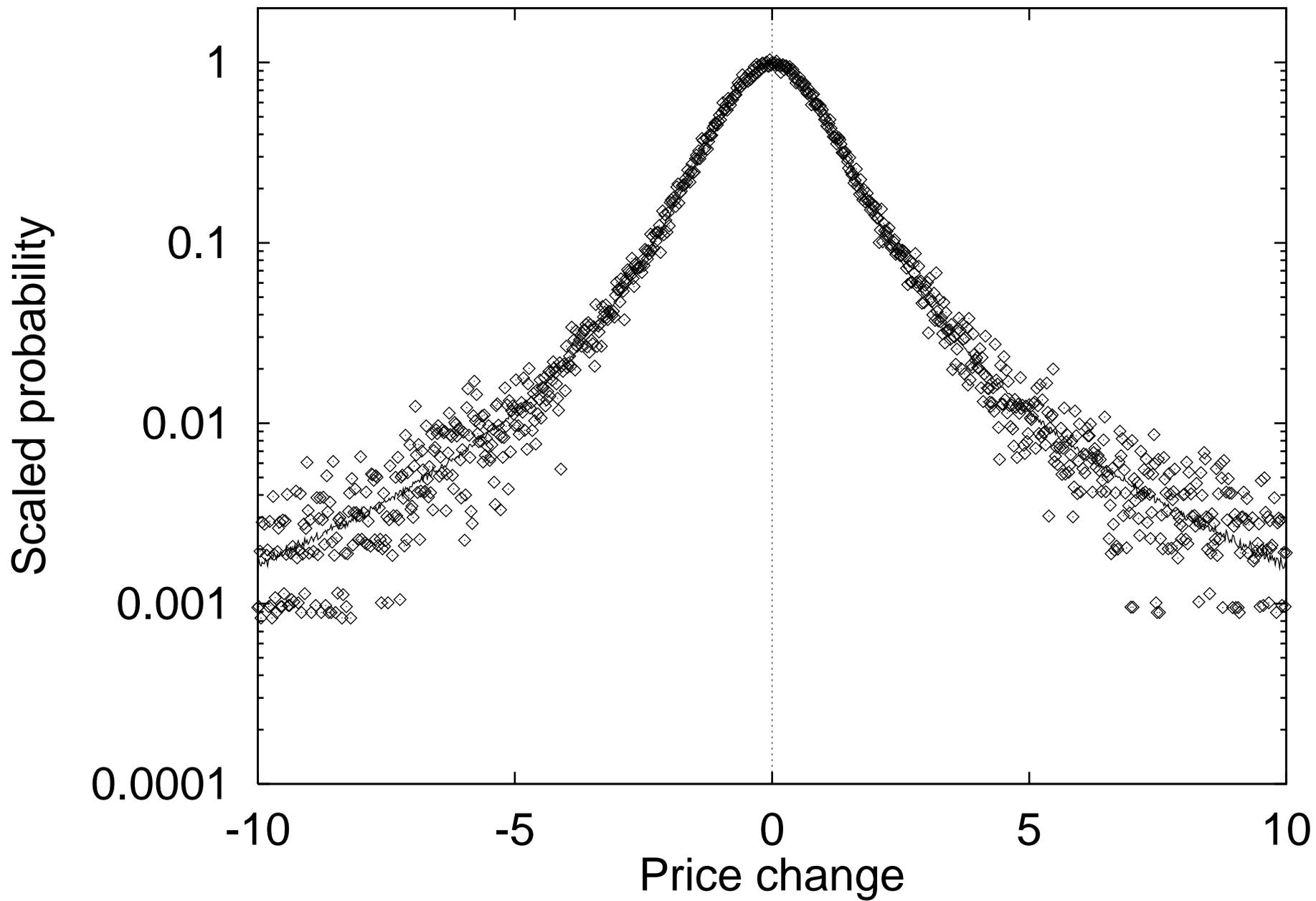
\noindent{\bf Fig.2:} The continuous line corresponds to our 
simplified model before switching on the temperature and local 
fields. The symbol $\diamond$ corresponds to a situation 
where all the super-spins, subjected to time-independent random 
local fields of magnitude $N J$, are maintained at a fictituous 
$T (\gg NJ)$ so that the time-averaged $M$ vanishes. Both 
the height and width of the distributions have been scaled to unity. 
The common parameters are $N = 10^3$ and $a = 0.05$.

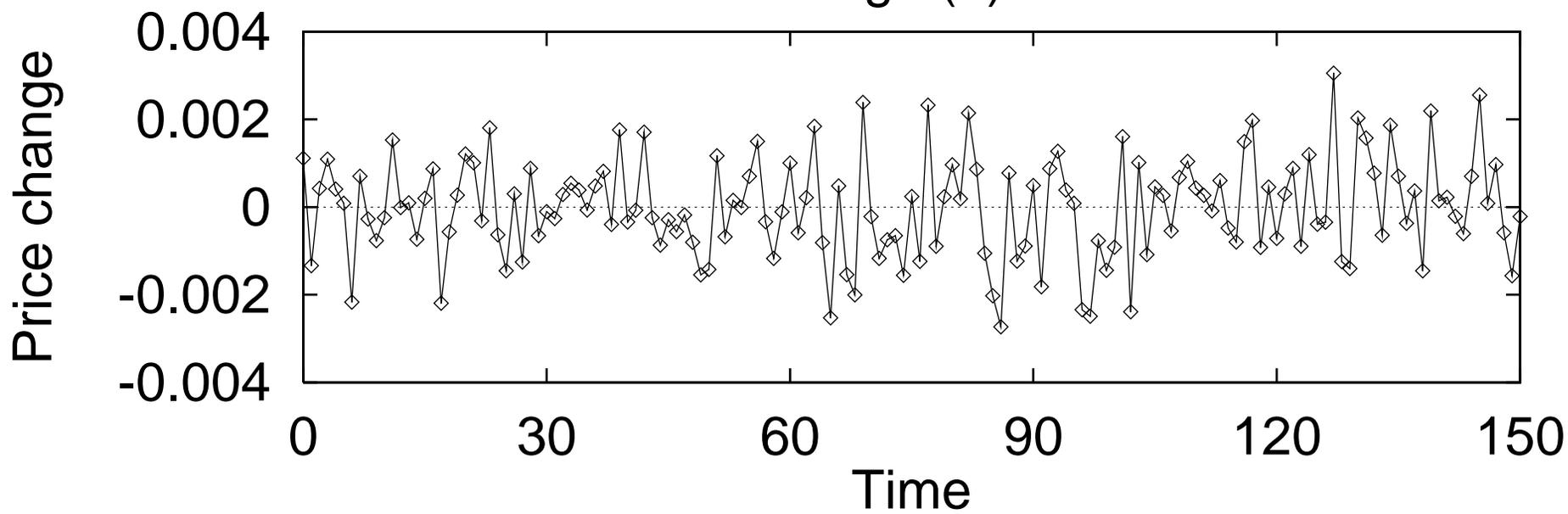
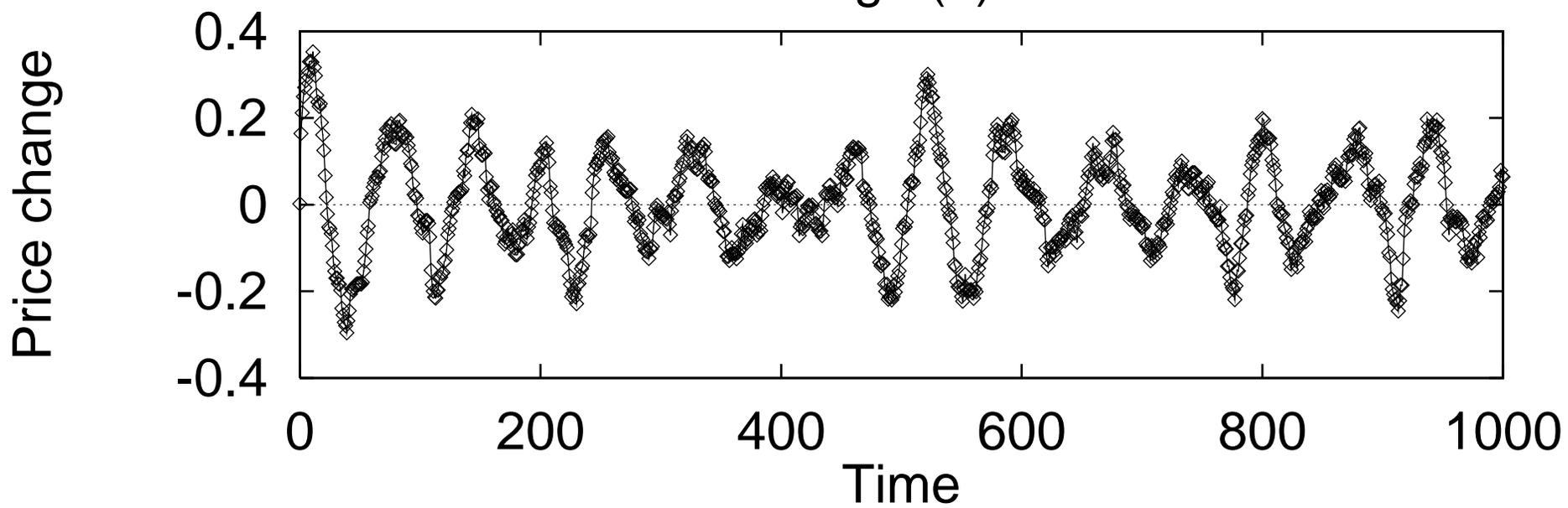
\noindent{\bf Fig.3:}\\
The fluctuations of the price changes (a) in the CB model 
and (b) in  a situation where all of the randomly chosen 
$50 \%$ of the super-spins are subjected to time-dependent local 
fields (whose magnitude is much larger than $T$ but sign is same 
everywhere) which flip when price change per individual investor 
goes out of the window $-0.4 \leq M \leq 0.4$.

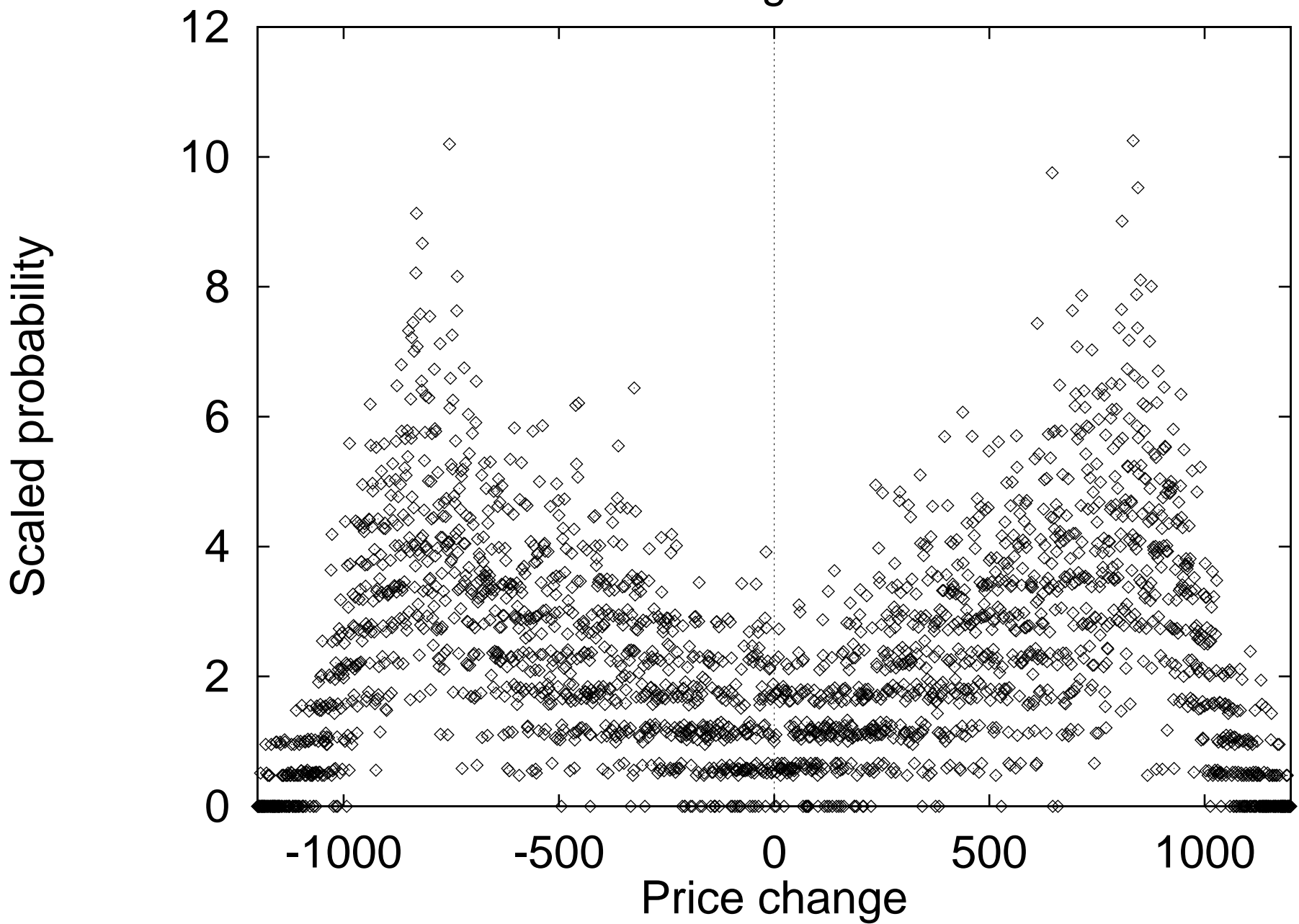
\noindent{\bf Fig.4:} The distribution of the price changes, 
corresponding to the the situation in fig.3(b), are shown, after 
scaling the probability for zero price change to unity. 


\newpage 

\begin{thebibliography}{999}

\bibitem{pareto} V. Pareto, {\it Cours d'Economique Politique}, vol.2; 
reprinted in {\it Oevres Completes de Vilfredo Pareto}, I. edited by 
G.H. Bousquet and G. Busino (Libraire Droz, Geneve, 1964).

\bibitem{bachelier} L. Bachelier {\it Theorie de la Speculation} 
(Gauthier-Villars, Paris, 1900)

\bibitem{mandel} B. Mandelbrot,  Comptes Rend. Acad. Sci. (Paris) {\bf 249},
313 (1959), reprinted in B. B. Mandelbrot, {\it Fractales, Hasard et Finance}. 
Flammarion, Paris 1997; B. Mandelbrot, J. Business Univ. Chicago {\bf 36}, 
394 (1963), {\bf 39}, 242 (1966) and  {\bf 40}, 393 (1967).

\bibitem{solomon} M. Levy and S. Solomon, Int. J. Mod. Phys. C {\bf 7}, 595 
(1996)

\bibitem{orlean} A. Orlean, La Recherche {\bf 22}, 668 (1991) 

\bibitem{anderson} P.W. Anderson, K.J. Arrow and D. Pines (eds.) 
{\it The Economy as an Evolving Complex System} (Addison-Wesley, 1988) 

\bibitem{arthur} W.B. Arthur, S. Durlauf and D. Lane {\it The Economy as 
an Evolving Complex System II} (Addison-Wesley, 1997) 

\bibitem{boupot} J.P. Bouchaud and M. Potters, {\it Theorie des Risques 
Financieres} (Alea-Saclay/Eyrolles, 1997);  R.N. Mantegna and H.E. Stanley, 
Nature {\bf 376}, 46 (1995); P. Gopikrishnan, M. Meyer, L.A. Nunes Amaral and 
H.E. Stanley, Eur. Phys. J. B {\bf 3}, 139 (1998).

\bibitem{kerkon} J. Kertesz and I. Kondor {\it Econophysics: An Emerging 
Science} (Kluwer, Boston, 1998)

\bibitem{lls} M. Levy, H. Levy and S. Solomon, J. de Physique I {\bf 5}, 1087 
(1995); see also M. Levy and S. Solomn, Physica A {\bf 242}, 90 (1997) and 
references therein. 

\bibitem{bak} P. Bak, M. Paczuski and M. Shubik, Physica A {\bf 246}, 430 (1997)

\bibitem{sornet} D. Sornette and A. Johansen, Physica A (in press); A. Johansen,
O. Ledoit, and D. Sornette, cond-mat/9810071. 

\bibitem{cb} R. Cont and J.P. Bouchaud, preprint cond-mat/9712318 and
page 71 in Bouchaud and Potters, ref. \cite{boupot}

\bibitem{stauaha} D. Stauffer and A. Aharony, 
{\it Introduction to Percolation Theory},
Taylor and Francis, London 1994; A. Bunde and S. Havlin, {\it Fractals and
Disordered Systems}, Springer, Berlin-Heidelberg 1996; M. Sahimi, {\it
Applications of Percolation Theory}, Taylor and Francis, London 1994.

\bibitem{staupen} D. Stauffer and T.J.P. Penna, Physica A {\bf 256}, 284 (1998);
D. Stauffer, P.M. C. de Oliveira and A.T. Bernardes, Int. J. Theor. Appl. 
Finance, in press; I. Chang and D. Stauffer, Physica A, in press.

\bibitem{stauap} D. Stauffer, Ann. Physik (Leipzig), Nov. 1998,

%\bibitem{cs} D. Chowdhury and D. Stauffer, unpublished.

\bibitem{weiss}G.H. Weiss, {\it Aspects and Applications of the Random Walk} 
(North-Holland, Amsterdam, 1994) 

\bibitem{hughes} B.D. Hughes {\it Random Walks and Random Environments}, 
(Oxford University Press, Oxford, 1995), vol.1

\bibitem{sornlevy} We are indebted to D. Sornette for this analysis, eqs.(3-10).

\bibitem{lux}  T. Lux, Economic J. {\bf 105}, 881 (1995);
 T.Lux and M. Marchesi, preprint.

\bibitem{mark} G.W. Kim and H.M. Markowitz,  J. Portfolio Management, Fall 1989,
45 (1989)

\end{thebibliography}
\end{document}